\documentclass{article}

\usepackage[utf8]{inputenc}
 \usepackage{graphicx}
\usepackage{amsmath}
\usepackage{array}
\usepackage{upgreek}
 \DeclareUnicodeCharacter{03B2}{\ensuremath{\beta}}
 \DeclareUnicodeCharacter{03B5}{\ensuremath{\epsilon}}
 \DeclareUnicodeCharacter{2212}{\ensuremath{-}}
 \usepackage[main, final]{causcien}
\workshoptitle{CauScien}

\usepackage[utf8]{inputenc} % allow utf-8 input
\usepackage[T1]{fontenc}    % use 8-bit T1 fonts
\usepackage{hyperref}       % hyperlinks
\usepackage{url}            % simple URL typesetting
\usepackage{booktabs}       % professional-quality tables
\usepackage{amsfonts}       % blackboard math symbols
\usepackage{nicefrac}       % compact symbols for 1/2, etc.
\usepackage{microtype}      % microtypography
\usepackage{xcolor}         % colors
\usepackage{natbib} % for bibliography
\setcitestyle{numbers}
\setcitestyle{square}
\usepackage{multirow}

\newcommand{\beginsupplement}{%
        \setcounter{table}{0}
        \renewcommand{\thetable}{S\arabic{table}}%
        \setcounter{figure}{0}
        \renewcommand{\thefigure}{S\arabic{figure}}%
     }

\title{Dynamic causal discovery in Alzheimer's disease through latent pseudotime modelling}

\author{Natalia Glazman \\
King's College London \\
\texttt{natalia.glazman@kcl.ac.uk}
  \And
  Jyoti Mangal \\
  King's College London \\
  \texttt{jyoti.mangal@kcl.ac.uk} \\
  \AND
  Pedro Borges \\
  Hologen \& King's College London \\
  \texttt{pedro.borges@hologen.ai} \\
  \And
  Sebastien Ourselin
  \\
  King's College London \\
  \texttt{sebastien.ourselin@kcl.ac.uk} \\
  \And
M. Jorge Cardoso \\
  King's College London \\
  \texttt{m.jorge.cardoso@kcl.ac.uk} \\
}

\begin{document}

\maketitle

\begin{abstract}

The application of causal discovery to diseases like Alzheimer's (AD) is limited by the static graph assumptions of most methods; such models cannot account for an evolving pathophysiology, modulated by a latent disease pseudotime. We propose to apply an existing latent variable model to real-world AD data, inferring a pseudotime that orders patients along a data-driven disease trajectory independent of chronological age, then learning how causal relationships evolve. Pseudotime outperformed age in predicting diagnosis (AUC 0.82 vs 0.59). Incorporating minimal, disease-agnostic background knowledge substantially improved graph accuracy and orientation. Our framework reveals dynamic interactions between novel (NfL, GFAP) and established AD markers, enabling practical causal discovery despite violated assumptions.
\end{abstract}

\section{Introduction}
\label{headings}
With nearly \$380 billion allocated annually to Alzheimer's disease (AD) research \cite{noauthor_report_2025}, clinical trials continue failing to halt disease progression \cite{kim_alzheimers_2022}. This ongoing challenge arises from the complexity of the disease, involving thousands of pathways whose interactions remain poorly understood \cite{arnold_pathways_2024}. Causal inference offers a powerful framework for modelling these relationships. For example, causal graph discovery methods can recover causal graphs from observational data, enabling identification and estimation of causal effects. Mapping AD's underlying causal relationships could yield more effective treatments, facilitate biomarker discovery, and enable personalised treatment plans.

Applying causal discovery to healthcare data presents distinct challenges. First, the absence of empirical ground truth hinders development and validation of domain-specific tools. Second, fundamental assumptions of causal discovery, such as absence of cyclical graphs and unobserved confounders, are frequently violated or untestable. While recent methodological advances allow analysis of more realistic data by relaxing these assumptions \cite{forre_constraint-based_2018, richardson_discovery_2013, huang_causal_2020}, practical applications remain limited. Current AD studies predominantly focus on static consensus graphs, neglecting dynamic relationships and latent variables that characterise the disease \cite{shen_challenges_2020}. Developing robust models of causal relationships in AD will advance both disease understanding and clinical application.

%as causal discovery tools move towards the relaxation of rigid assumptions about the data and underlying causal mechanisms, allowing us to move closer to the clinical application of these methods. 

%It is important to establish a known graph to test these models against 

%It has been shown that many disease-related factors, including some of the key drivers of disease, have differing effects throughout the development of disease; for example, at early stages or pre-clinical stages of AD, tau protein phosphorylation has an important role in maintaining cognitive function \cite{guo_roles_2017}. 

Disease progression rates vary across AD patients \cite{stern_influence_1994}, partly reflecting "cognitive reserve" - protective brain mechanisms arising from poorly characterised biological, lifestyle, and genetic factors \cite{stern_cognitive_2012, nelson_cognitive_2021}. While approximating cognitive reserve through education and socioeconomic status partially explains individual variation, explicitly modelling latent disease pathway modulators remains crucial for comprehensive understanding of AD. The abstraction of pseudotime, which is a latent (unobserved) dimension measuring the progress of cell states through a transition is frequently applied to molecular-level disease modelling, and recent research has adopted this abstraction (both categorical and continuous) for modelling disease progression using electronic health records \cite{heumos_open-source_2024, young_uncovering_2018}.

Here, we apply causal discovery to real-world AD data through a latent pseudotime model introduced by \citet{zhou_individualized_2023}, investigating how causal interactions evolve along disease progression. We leverage known causal relationships from literature and validate our model through qualitative and quantitative evaluation against a consensus graph. Our analysis incorporates key AD biomarkers to characterise their interactions, including plasma NfL and GFAP, recently emerged as novel biomarkers for AD detection \cite{wang_peripheral_2024, ingannato_plasma_2024, bolsewig_association_2024}.

\section{Methods}
\label{headings}

\textbf{Data} We analysed data from the Alzheimer's Disease Neuroimaging Initiative (ADNI) dataset \cite{petersen_alzheimers_2010}, selecting 16 variables with established or putative causal roles in AD aetiology. These comprised demographic variables (years of education, sex, age, and APOE genotype), segmented brain region volumetric measurements (total intracranial volume (ICV), hippocampus, amygdala, temporal lobe), plasma biomarkers (A$\upbeta$40, A$\upbeta$42, pTau217, NfL, GFAP), and cognitive scores (Trail Making Test Part B (TRABSCORE), ADAS-Cog-13 (ADAS13), Rey Auditory Verbal Learning Test immediate (RAVLT)). The sample consisted of the following participant numbers, broken down by diagnosis: 48 AD, 117 MCI (mild cognitive impairment), and 215 CN (cognitively normal). \\

% The model’s application rests on both theoretical and empirical grounds. Theoretically, the pseudotime variable is identifiable up to monotonic transformations if causal effects vary along its trajectory, as shown by \citet{zhou_emerging_2023}. Empirically, the inferred pseudotime stratifies patients by diagnosis, predicts disease status better than age, and captures causal dynamics consistent with known AD pathology, demonstrating that it serves as a robust and clinically meaningful proxy for true disease progression.

\textbf{Pseudotime model of Alzheimer's disease} We modelled AD progression using cross-sectional data from the ADNI dataset with Bayesian Networks with Latent Time Embedding (BN-LTE), a model proposed by \citet{zhou_individualized_2023} to estimate causal graphs as a function of pseudotime. In disease modelling, pseudotime represents a latent variable that orders samples along a continuous trajectory of disease progression, capturing relative disease states rather than chronological time \cite{he_multi-modal_2025, hong_image-level_2022}. This enables modelling of dynamic changes in biomarkers or molecular states when using irregularly sampled or cross-sectional data.

Theoretical identifiability of the pseudotime variable up to monotonic transformations is shown by \citet{zhou_individualized_2023} under the condition that causal relationships change along the pseudotime axis.

 %As opposed to the original paper, we can pose this problem through the lens of a latent confounder that may represent cognitive reserve, and evaluate the validity of the causal graph output.

We implemented the framework of \citet{zhou_individualized_2023}, described as follows. We assume faithfulness and causal sufficiency when incorporating the latent pseudotime variable. Let $\mathbf X = (X_1,..., X_p) \in \mathbb{R}^p$ be a p-dimensional random variable vector, and let $G= (V,E)$ denote a directed acyclic graph (DAG) consisting of a node set $V = \{1,...,p\}$ corresponding to elements of $\textbf{X}$, and edge set $E \subset V \times V$ representing causal relationships between them. The distribution of $\textbf{X}$ factorizes as \begin{equation}
    F(\textbf{X}|Z,\Theta) = \prod^p_{j=1} F_j(X_j|X_{l \in pa_{G(Z)}}(j) , Z, \Theta),
\end{equation} 
where $Z \in \mathbb{R}^n$ denotes pseudotime, $pa_{G(Z)}(j)$ denotes the parent nodes of $j$, and $\Theta$ represents model parameters. The conditional probability distribution of a single biomarker, $F_j$, is modelled as 
\begin{equation}
    X_{j} = a_j(Z) +\sum^p_{l=1} b_{jl}(Z) X_{l} + \epsilon_j, \epsilon_j \sim N(0,\sigma_j^2),
\end{equation} 
where $a_j(Z)$ represents the baseline function governing biomarker progression across pseudotime, $b_{jl}(Z)$ captures pseudotime-dependent causal effects, and $\epsilon_j$ denotes per-variable noise. Both functions are parametrised as cubic b-splines, following \citet{zhou_individualized_2023}. Parameter posteriors were obtained using Markov chain Monte Carlo (MCMC). Details of the priors, sampling, posterior estimation, and convergence statistics are provided in the Supplementary Material. 

%Forbidden edges were skipped during sampling of causal edges, $b_{jl}$ and static variables were skipped during sampling of the spline coefficients corresponding to the baseline pseudotime dependent term, $a_j(z)$.

\textbf{Background knowledge} Background or expert knowledge is commonly used to improve causal graph discovery performance when model assumptions are violated \cite{sinha_using_2021, borboudakis_incorporating_2012, castelo_priors_2000}. We introduced disease-agnostic background knowledge to enhance model performance. To minimize bias, our background knowledge exclusively constrained causal and pseudotime relationships directed towards immutable variables (e.g., sex, genotype – considered root nodes) and prohibited cognitive score variables from forming outgoing causal effects (sink nodes). Although cognitive variables may influence lifestyle choices, such effects are unlikely to significantly impact other variables in the elderly ADNI population. Additional implementation details are found in the Supplementary Material.

\section{Results}
\label{headings}

\begin{figure}[h!]
    \centering\includegraphics[width=12cm]{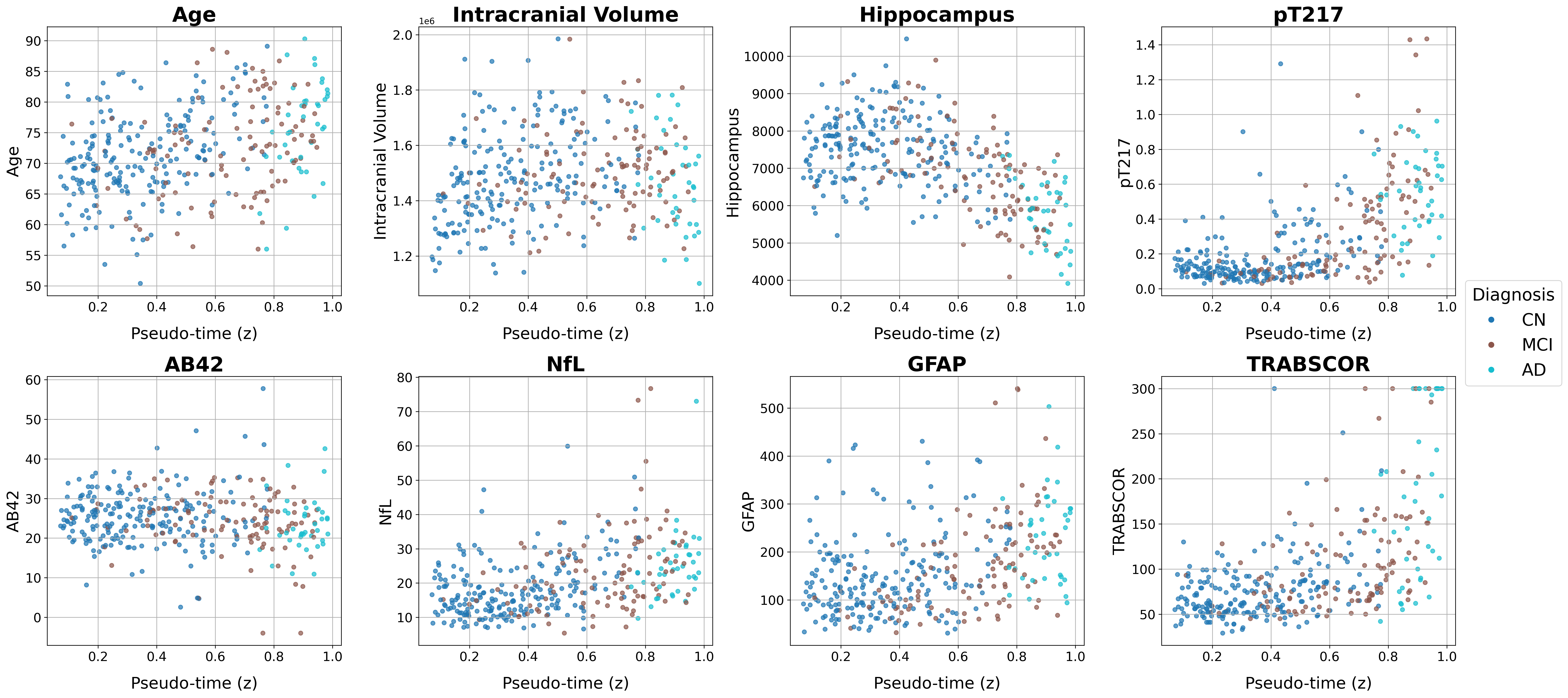}
    \caption{Demographic variables and biomarkers plotted as a function of the posterior mean of the inferred pseudotime values (z) of a single chain inferred by the BN-LTE model, with patients colour-coded according to their diagnosis.}
    \label{fig:memory_allen}
\end{figure}

The patient distribution along the pseudotime axis corresponded to disease severity (Fig. \ref{fig:memory_allen}), with cognitively normal participants clustering at early pseudotime, MCI patients concentrated in the intermediate region, and AD patients positioned towards late pseudotime. This ordering aligned with expected AD-associated biomarker trajectories, including decreased hippocampal volume, deteriorating cognitive scores, and elevated NfL and GFAP levels. To verify that pseudotime captured disease progression rather than chronological ageing, we compared predictive performance using a logistic generalized additive model, which captures complex, non-linear relationships. Pseudotime demonstrated superior predictive performance for diagnosis, achieving a mean AUC of 0.82 (95\% CI: 0.81, 0.82) (p < 0.001) compared to 0.59 (p < 0.01) for age alone. We identified edges with the highest posterior inclusion probability throughout pseudotime for edges between variables with no background knowledge across different settings and validated them against literature evidence (Table \ref{tab:table1}). The model correctly identified several known disease-related pathways, including the causal effect of plasma NfL on hippocampal volume reduction and pTau-induced NfL elevation \cite{gaetani_neurofilament_2019, didonna_tau_2020}.

\begin{table}[h]
\footnotesize
\begin{tabular}{ >{\centering\arraybackslash}p{2.8cm} >{\centering\arraybackslash}p{1.9cm} >{\centering\arraybackslash}p{1.7cm} >{\centering\arraybackslash}p{2cm} >{\centering\arraybackslash}p{2.9cm} }
\hline
%    \multirow{2}{*}{\textbf{Edge}} &
%      \multicolumn{3}{c}{\textbf{PIP}} & \\
%\cline{2-4}
\textbf{Edge}  & \textbf{PIP (No mod.)} & \textbf{PIP (RN)} &  \textbf{PIP (RN + SN)} & \textbf{Lit. consensus} \\
 \hline
 pTau217 $\rightarrow$ GFAP   & 0.64 (0.38)    & 0.93 (0.13) & 0.80 (0.16) & Possible/unknown \cite{didonna_tau_2020}  \\
   \hline
A$\upbeta$42 $\rightarrow$ A$\upbeta$40  & 0.75 (0.04)  & 0.75 (0.43) & 0.75 (0.43) & Present \cite{tran_cross-seeding_2017} \\
 \hline
pTau217 $\rightarrow$ NfL   & 0.35 (0.27) & 0.61 (0.17) & 0.57 (0.15) & Possible \cite{didonna_tau_2020}   \\
 \hline
NfL $\rightarrow$ Hippocampus   & 0.25  (0.18)  & 0.63 (0.01)  & 0.53 (0.18) & Possible \cite{jung_potential_2023} \\
 \hline
A$\upbeta$42 $\rightarrow$ NfL  & 0.60 (0.07)   & 0.49 (0.12) & 0.46 (0.07) & Possible \cite{zhang_amyloid_2023} \\
 \hline
\end{tabular}\\
\caption{\label{tab:table1}Causal edges identified using BN-LTE with background knowledge, with the posterior probability of inclusion (PIP) junder different background knowledge settings. Values shown are chain-level mean and standard deviation for modifications of root nodes (RN) and sink nodes (SN).}
\end{table} 

To comprehensively evaluate model performance, we compared our estimated graphs against a consensus graph constructed from Alzheimer's literature (Fig. \ref{fig:graphs}; further details listed in the Supplementary Material). The baseline model without modifications performed poorly across both edge detection and directionality metrics. Performance improved substantially with the incorporation of background knowledge and static features (Table \ref{tab:table2}). The improvement was particularly pronounced in orientation metrics and was reflected in the reconstruction accuracy.

After incorporating background knowledge for root and sink nodes, we evaluated performance improvements for variables without background knowledge constraints - specifically, biomarkers and brain volumetric measures. Given the induced orientation bias, we assessed this graph subset based only on edge presence rather than directionality. The model with background knowledge outperformed the unmodified model across various metrics (Table \ref{tab:table2}).

\begin{figure}[h!]
    \centering
    \includegraphics[width=0.75\linewidth]{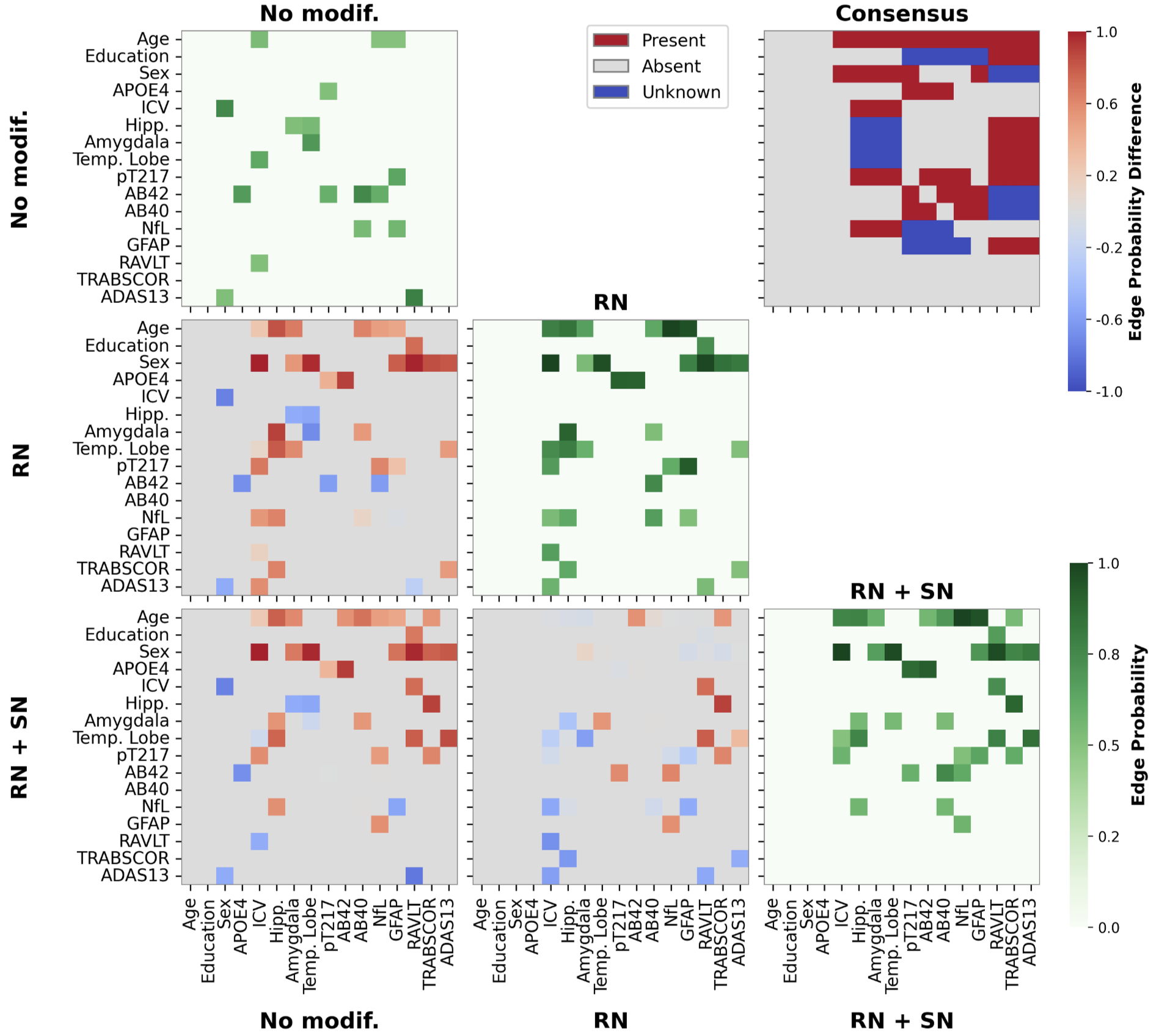}
    \caption{Established consensus graph compared to estimated causal graphs presented as adjacency matrices with edge indices averaged across pseudotime and over the four chains, resulting in an edge inclusion probability (only PIP > 0.5 shown). The matrices on the diagonal reflect the resulting matrices from different settings, while the strictly lower triangular matrices reflect differences between each pair of matrices.}
    \label{fig:graphs}
\end{figure}

\begin{table}[h!]
\footnotesize
\begin{tabular}{ >{\centering\arraybackslash}m{0.8cm} >{\centering\arraybackslash}m{1.1cm} >{\centering\arraybackslash}m{1.1cm} >{\centering\arraybackslash}m{1.1cm} >{\centering\arraybackslash}m{1.1cm} >
{\centering\arraybackslash}m{1.1cm} >
{\centering\arraybackslash}m{1.1cm} >
{\centering\arraybackslash}m{0.7cm} >
{\centering\arraybackslash}m{0.7cm} >
{\centering\arraybackslash}m{0.8cm}}
\hline
    \multirow{2}{*}{\textbf{Modif.}} &
      \multicolumn{2}{c}{\textbf{Presence}} &
      \multicolumn{2}{c}{\textbf{Orientation}}  & 
      %\multirow{2}{*}{\parbox{1.4cm}{Subset precision}} &
      \multicolumn{2}{c}{\textbf{Subset (Presence)}}  & 
      \multirow{2}{*}{\textbf{MSE}} & 
      \multirow{2}{*}{\textbf{SHD}} & 
      \multirow{2}{*}{\parbox{0.8cm}{\textbf{Subset SHD}}}\\
\cline{2-7}
  & P & R & P & R & P & R & & &\\
 \hline
None &  0.80 \scriptsize{[0.80,82]}  & 0.16 \scriptsize{[0.16,16]} &  0.62 \scriptsize{[0.62,0.63]} & 0.50 \scriptsize{[0.49,0.50]} & 0.67 \scriptsize{[0.65,0.71]} & 0.10 \scriptsize{[0.10,0.10]} & 0.98 \scriptsize{(0.045)} & 67 \scriptsize{[67,67]} & 21 \scriptsize{[21,21]}\\
 \hline
RN  & 0.72 \scriptsize{[0.70,0.74]}  & 0.35 \scriptsize{[0.33,0.36]} & 0.89 \scriptsize{[0.89,0.90]} & 0.84 \scriptsize{[0.84,0.84]} & 0.74 \scriptsize{[0.72,0.74]} & 0.29 \scriptsize{[0.29,0.30]} & 0.95 \scriptsize{(0.016)} & 53 \scriptsize{[55,53]} & 23 \scriptsize{[24,23]} \\
 \hline
RN + SN  &  0.88 \scriptsize{[0.88,0.89]}  & 0.45 \scriptsize{[0.42,0.45]} & 0.96 \scriptsize{[0.96,0.96]}  & 0.88 \scriptsize{[0.88,0.88]} & 0.79 \scriptsize{[0.79,0.79]} & 0.39 \scriptsize{[0.38,0.39]} & 0.93 \scriptsize{(0.025)} & 41 \scriptsize{[43,41]} & 21 \scriptsize{[21,21]} \\
 \hline
\end{tabular}\\
\caption{\label{tab:table2}Statistics of causal graphs estimated with BN-LTE. Presence precision (P) and recall (R) measure edge presence using symmetric adjacency matrices. Orientation metrics evaluate directionality for edges present in both consensus and estimated graphs. Subset metrics exclude variables with background knowledge. Values presented as median, [95\% CI] apart from MSE (SD).}
\end{table}

%We identify edges that are unlikely to be present as they do not align with the AT(X)N disease cascade. This framework categorizes the disease status of AD patients according to their biomarker and symptom status, which results in a consensus biomarker cascade of the following order: amyloid beta dysfunction, tau hyperphosphorylation, intermediate biomarker changes, neurodegeneration \cite{hampel_developing_2021}. Therefore, it is more unlikely that, for example, that neurodegeneration has a causal effect on the amyloid pathway, discounting feedback loops. 

\begin{figure}[h]
        \centering\includegraphics[width=13cm]{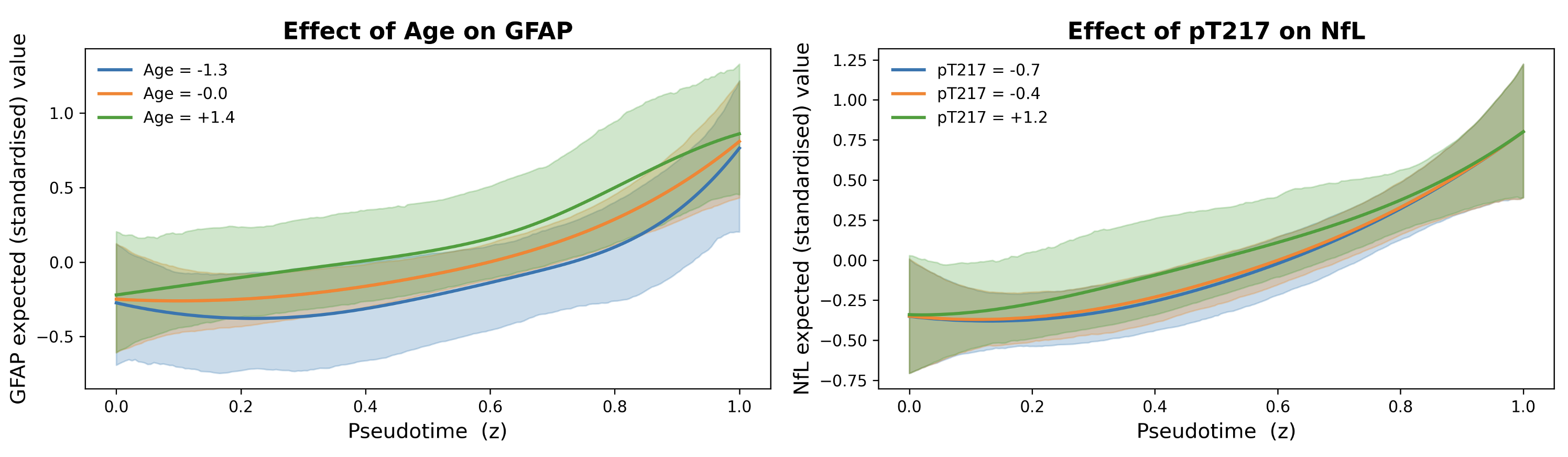}
    \caption{Biomarkers plotted against pseudotime, with solid lines representing the mean baseline trajectory of biomarkers and the corresponding effect of changing a specific parent variable. Parent variables are changed to the 5th, 50th, and 95th percentiles of the standardized values. The solid lines correspond to the mean of posteriors with shaded areas being 95\% CI.}
    \label{fig:changing}
\end{figure}

The causal relationships between variables exhibited dynamic changes across pseudotime; for example, age showed a constant effect on GFAP, while pTau's influence on NfL manifested early in pseudotime progression (Fig \ref{fig:changing})\cite{hampel_developing_2021}. This pattern is consistent with AD pathophysiology, where the effect of pTau is known to plateau, while the effect of age on neuroinflammation most likely remains constant over the course of the disease \cite{bertsch_role_2023}.

%This aligns with our understanding of AD disease progression, modelled by the AT(X)N cascade. This framework categorizes the disease status of AD patients according to their biomarker and symptom status, which results in a consensus biomarker cascade of the following order: AB dysfunction, tau hyperphosphorylation, intermediate biomarker changes, neurodegeneration 
We also identified edges that contradict established literature; for example, the proposed causal effect of pTau $\rightarrow$ GFAP and NfL $\rightarrow$ A$\upbeta$40 (Fig. \ref{fig:graphs}) contradict studies that show that amyloid dysfunction precedes neurodegeneration, and GFAP elevation precedes changes in pTau \cite{bilgel_longitudinal_2023}.
\section{Discussion}
Our findings demonstrate significant progress toward clinically applicable causal discovery in AD. Using BN-LTE, we estimated causal graphs as a function of pseudotime. Importantly, incorporating background knowledge proved essential for generating reliable causal graphs when analysing real-world data where model assumptions are violated. Notably, this knowledge remained disease-agnostic, constraining only demographic variables and cognitive outcomes, which significantly improved the recovered graphs. We identified both known and novel causal dynamics along the latent pseudotime through incorporation of emerging biomarkers (NfL and GFAP).

This pseudotime framework offers immediate translational potential. It can revolutionize clinical trial design by precisely stratifying patients for targeted interventions at optimal disease stages, potentially explaining heterogeneous treatment responses and identifying therapeutic windows. Furthermore, dynamic causal relationships across pseudotime suggest that combination therapies may require tailored sequencing based on a patient's disease stage.

AD progression arises from a complex interplay of unobserved factors such as cognitive reserve and lifestyle-driven neuroprotection, alongside other modulators of its pathological pathways \cite{zhang_recent_2024}. Under the framework of \cite{huang_causal_2020}, our approach treats pseudotime as a surrogate for latent modulating factors that drive non-stationary causal mechanisms throughout disease progression, ensuring identifiability of the causal graph. The inferred pseudotime separates patients by disease stage, predicts disease status better than age, and captures causal dynamics consistent with known AD pathology, demonstrating that it serves as a robust and clinically meaningful proxy for true disease progression. While this work demonstrates the value of a one-dimensional abstraction, future work could model these drivers as a multivariate construct to achieve a more granular understanding of AD’s heterogeneity.

Several methodological advances would strengthen clinical applicability. First, model assumptions must be relaxed to accommodate biomedical systems with unobserved confounders. Second, our analysis was limited by dataset size and variable selection, preventing the inclusion of factors like ethnicity due to insufficient cohort heterogeneity. Expanding variables and sample size should enhance causal graph accuracy. Recent multi-dataset causal discovery advances \cite{mooij_joint_2020} could enable more robust causal identification through cross-validation across independent cohorts. In addition, an evaluation of this model using longitudinal data could both validate the model and provide insights on AD progression through predictions grounded in causal reasoning.

In summary, our work establishes a foundation for integrating causal discovery into AD research and clinical practice. Combining pseudo-time modelling with disease-agnostic background knowledge provides a practical framework for balancing theoretical rigour with clinical applicability. As we refine these methods and expand to larger, diverse datasets, causal discovery can transform our understanding of AD heterogeneity and guide precision medicine development.

\paragraph{Acknowledgments} Natalia Glazman is supported by the King's College London Digital Twins for Healthcare CDT and cofunded by Siemens Healthineers. Jyoti Mangal is supported by NHS England.

\newpage

\bibliographystyle{abbrvnat}

{\small
\bibliography{references.bib}}

\newpage

\section{Supplementary material}
\beginsupplement

\subsection{MCMC procedure and convergence statistics}

In order to confirm the stability of the MCMC results, we ran 4 chains across 5000 iterations with 1000 iterations for burn-in and evaluate their convergence statistics. We used 5 knots for both causal and baseline cubic b-splines. The rest of the hyperparameters were set according to \citet{zhou_individualized_2023}. The \^R values and ESS values are presented in the table below.

\begin{table}[ht]
\centering
\begin{tabular}{ >{\centering\arraybackslash}p{3.2cm} >{\centering\arraybackslash}p{1cm} >{\centering\arraybackslash}p{1cm} >{\centering\arraybackslash}p{1cm} >{\centering\arraybackslash}p{1cm} >{\centering\arraybackslash}p{1cm} >{\centering\arraybackslash}p{1cm} }
 \\[-1em]
 \hline
\multirow{2}{*}{\textbf{Variable}} & \multicolumn{2}{c}{\textbf{No mod.}}  &
\multicolumn{2}{c}{\textbf{SF}} &
\multicolumn{2}{c}{\textbf{SF + BK}} \\
 \cline{2-7}
  &  \^R &  ESS & \^R & ESS & \^R &  ESS \\
\\[-1em]
 \hline
Per-variable noise variance   & 1.02  & 1269.58 & 1.03 & 364.62 & 1.01   & 400.10  \\
 \hline
Causal spline coefficients  & 1.00  & 4041.07 & 1.00 & 730.3 & 1.00   & 723.14  \\
 \hline
Baseline spline coefficients & 1.03  & 346.7 & 1.01 & 432.10 & 1.01   & 217.42  \\
 \hline
Roughness parameter & 1.00 & 1504.03 & 1.00 & 711.52 & 1.01  & 250.85 \\
 \hline
Variance of spline coefficients & 1.01& 233.74 & 1.01 & 219.91 & 1.01    & 199.33  \\
 \hline
\end{tabular}\\
\caption{Convergence statistics on the MCMC results}
\end{table} 

We followed the procedure outlined in the supplementary material of \citet{zhou_individualized_2023} in order to sample the parameter posteriors. To obtain the PIPs of the edges, we averaged the edge inclusion indicator parameters ($r_{jk}$) across the four chains, and used a threshold of $P\ge0.5$ to construct the final causal graphs, which were then used to calculate statistics in Table \ref{tab:table2}.

In addition, we remove the Coulomb prior as described by \citet{zhou_individualized_2023} from the model, which serves to ensure a distance between samples and prevent clustering, as we do not make an assumption that there exists a uniform distribution of patients across disease pseudotime due to the uneven distribution of patient diagnoses, as outlined in the Methods section.

All experiments were performed on a MacBook Pro with an Apple M4 Pro CPU and 24GB of RAM. A single run of the MCMC algorithm (4 chains) required approximately 30 minutes to complete.

Implementation of background knowledge involved specifying the variables that were root nodes and sink nodes. Root nodes were then skipped during sampling of baseline and incoming causal trajectory variables, and remained set at 0. For sink nodes, only outgoing causal trajectories were set to 0. When calculating residuals in the sampling, root nodes were excluded.

For settings with background knowledge, MSE was calculated over only the variables that were non-root nodes.

\subsection{Consensus and Estimated Causal Graph}
The consensus causal graph was created using a literature search of consensus mechanisms of AD progression. Where variables are grouped together, the citation shows evidence of the effect across all the variables. Below are the identified causal relationships:

Age → ICV\cite{caspi_changes_2020}, [Hippocampus, Amygdala, Temporal lobe]\cite{walhovd_consistent_2011}, pTau \cite{chatterjee_age-related_2023}, [A$\upbeta$42, A$\upbeta$40] \cite{zheng_healthcare_2020}, GFAP \cite{andronie-cioara_molecular_2023}, [RAVLT, TRABSCORE, ADAS13] 

Education → [RAVLT, TRABSCORE, ADAS13] \cite{stricker_mayo_2021}

Sex → [ICV, Hippocampus, Amygdala, Temporal lobe] \cite{ritchie_sex_2018}, pTau \cite{sundermann_sex_2020}, RAVLT, TRABSCORE, ADAS13 \cite{stricker_mayo_2021}

APOE → pTau\cite{ferrari-souza_apoe4_2023, young_apoe_2023, la_joie_association_2021, therriault_association_2020}, [A$\upbeta$42, A$\upbeta$40] \cite{kanekiyo_apoe_2014}

ICV → Hippocampus, Amygdala, Temporal lobe

Hippocampus → [RAVLT, TRABSCORE, ADAS13] \cite{langella_association_2021, botdorf_metaanalysis_2022}

Amygdala → [RAVLT, TRABSCORE, ADAS13] \cite{mcgaugh_involvement_1996, inman_direct_2018}

Temporal lobe → [RAVLT, TRABSCORE, ADAS13] \cite{ross_improved_2011, pobric_anterior_2007}

pTau → [A$\upbeta$42, A$\upbeta$40] \cite{bloom_amyloid-_2014}, NfL \cite{didonna_tau_2020}, [RAVLT, TRABSCORE, ADAS13] \cite{pereira_plasma_2021}

A$\upbeta$42 → A$\upbeta$40 \cite{hasegawa_interaction_1999}, pTau217 \cite{gillespie_testing_2025, wu_-amyloid_2018, bloom_amyloid-_2014}  NfL \cite{zhang_amyloid_2023}, GFAP \cite{hampel_amyloid-_2021} 

A$\upbeta$40 → A$\upbeta$42 \cite{hasegawa_interaction_1999}, pTau217 \cite{gillespie_testing_2025} 

NfL → [Hippocampus, Amygdala, Temporal lobe] \cite{gaetani_neurofilament_2019}

GFAP → [RAVLT, TRABSCORE, ADAS13] \cite{tan_peripheral_2023}

The rest of the edges were marked as being absent, aside from the following, which were left out of the analysis due to a lack of causal understanding in literature:

Education → [pTau217, A$\upbeta$40, A$\upbeta$42, NfL, GFAP]

Hippocampus - [Temporal Lobe, Amygdala]

[A$\upbeta$42, A$\upbeta$42] → [RAVLT, TRABSCORE, ADAS13]

[NfL, GFAP] → [pTau217, A$\upbeta$40, A$\upbeta$442]

GFAP → NfL

These edges were not included when calculating statistics such as precision, recall, and SHD. It is important to note that the consensus graph contains three feedback systems, namely between pTau217 and AB42 and AB40. This must be taken into account when considering the accuracy values obtained for graph recovery, as the model restricts any cyclical relationships.

\end{document}